\documentclass[runningheads]{llncs}

\usepackage[T1]{fontenc}

\usepackage{cite}
\usepackage{amsmath,amssymb,amsfonts}
\usepackage{algorithm}
\usepackage{algorithmic}
\usepackage{graphicx}
\usepackage{textcomp}
\usepackage{xcolor}
\usepackage{censor,caption}

\usepackage{tabularx}
\usepackage{multirow}
\usepackage{adjustbox}
\usepackage{booktabs}
\usepackage{multicol}

\usepackage{subcaption}

\def\BibTeX{{\rm B\kern-.05em{\sc i\kern-.025em b}\kern-.08em
    T\kern-.1667em\lower.7ex\hbox{E}\kern-.125emX}}

\begin{document}

\title{Development of an Edge Resilient ML Ensemble to Tolerate ICS Adversarial Attacks}
\titlerunning{Development of an Edge Resilient ML Ensemble}
%
\author{Likai Yao\inst{1} \and
Qinxuan Shi\inst{2} \and Zhanglong Yang\inst{2} \and Sicong Shao\inst{2} \and Salim Hariri\inst{1}}
%
%
\institute{NSF Center for Cloud and Autonomic Computing, University of Arizona, Tucson, AZ 85721 USA \and
School of Electrical Engineering and Computer Science, University of North Dakota, Grand Forks, ND 58202 USA
\\
}
%

\maketitle

\begin{abstract}
Deploying machine learning (ML) in dynamic data-driven applications systems (DDDAS) can improve the security of industrial control systems (ICS). However, ML-based DDDAS are vulnerable to adversarial attacks because adversaries can alter the input data slightly so that the ML models predict a different result.  In this paper, our goal is to build a resilient edge machine learning (reML) architecture that is designed to withstand adversarial attacks by performing Data Air Gap Transformation (DAGT) to anonymize data feature spaces using deep neural networks and randomize the ML models used for predictions. The reML is based on the Resilient DDDAS paradigm, Moving Target Defense (MTD) theory, and TinyML and is applied to combat adversarial attacks on ICS. Furthermore, the proposed approach is power-efficient and privacy-preserving and, therefore, can be deployed on power-constrained devices to enhance ICS security. This approach enables resilient ML inference at the edge by shifting the computation from the computing-intensive platforms to the resource-constrained edge devices. The incorporation of TinyML with TensorFlow Lite ensures efficient resource utilization and, consequently, makes reML suitable for deployment in various industrial control environments. Furthermore, the dynamic nature of reML, facilitated by the resilient DDDAS development environment, allows for continuous adaptation and improvement in response to emerging threats. Lastly, we evaluate our approach on an ICS dataset and demonstrate that reML provides a viable and effective solution for resilient ML inference at the edge devices.
\end{abstract}

\keywords{Edge AI  \and DDDAS \and Adversarial ML \and Cybersecurity}


\section{Introduction}
\vspace*{-0.4cm}

The rapid integration of Industrial Control Systems (ICS) into the broader network of interconnected devices has significantly improved operational efficiency across critical infrastructures such as power grids, manufacturing plants, and water treatment facilities. However, this increased connectivity also introduces substantial cybersecurity risks \cite{chen2020generating}. Deploying machine learning (ML) in dynamic data-driven applications systems (DDDAS) can improve the security of industrial control systems (ICS) \cite{darema2005grid}. By using ML and DDDAS techniques for ICS, we can dynamically adapt the detection systems to detect zero-day attack patterns that can be injected into large amounts of dynamic ICS data. However, according to recent studies, ML models can be fooled by adversarial examples~\cite{papernot2016limitations}. An adversary may modify the input data in a way that makes the results produced by the ML models differ from the expected results that the model is supposed to produce without the adversarial attack against the ML model.  As a result, ML models can make ICS resources and applications vulnerable to adversarial ML attacks \cite{yao2023resilient, chen2020generating}. On the other hand, traditional centralized ML approaches for DDDAS  can also be vulnerable due to latency and connectivity issues and require computationally intensive devices \cite{blasch2019study}. Exploring alternatives, it is important to use edge devices by pushing ML computations to the edge and the extreme edge of the network, i.e., run ML inference on microcontroller units (MCUs) and sensors, and thus that will reduce significantly the latency while improving performance and resilience. Thus, the goal of this paper is to develop a resilient edge machine learning (reML) architecture that leverages DDDAS paradigm~\cite{darema2005grid}, MTD theory~\cite{ge2014toward}, and TinyML~\cite{warden2019tinyml}.  The proposed reML is designed to withstand adversarial ML attacks and push ML inference to edge devices, allowing for efficient, low-power data analytics at the edge and the extreme edge of ICS networks.
\vspace*{-0.05cm}

The main advantages of reML are listed as follows: \textbf{DDDAS Paradigm:} reML supports the DDDAS paradigm by implementing a dynamic, adaptive, and resilient ML system that closely integrates data acquisition, model execution, and system control in a feedback loop, particularly focused on power-constrained edge AI environments and critical infrastructure applications; \textbf{On-device ML Inference}: reML can perform on-device ML inference for attack detection to provide real-time alerts and to identify potential security breaches as they occur; \textbf{Reduced Latency and Enhanced Privacy}: reML uses edge ML to process ICS data locally, thereby reducing the need for constant communication with central servers. This will enhance response times, and by keeping data on the edge,  data privacy can be improved. This is crucial for ICS, where rapid decision-making is essential to maintain operational integrity and security; \textbf{Resilience to Adversarial Attacks}: reML leverages MTD, which dynamically changes the system configuration by randomizing the selection of ML services at run-time and using Data Air Gap Transformation (DAGT) to anonymize data feature spaces, making it very difficult for adversaries to compromise the ML models, thus ensuring that attackers cannot effectively manipulate the ML models used to secure ICS operations; \textbf{Efficient Resource Utilization}: reML applies TinyML to optimize power consumption for low-power devices, and that will make reML ideal for industrial control environments that use edge devices and IoT resources. DAGT leverages quantized deep autoencoder neural networks to mask the data feature space on low-power edge and extreme edge devices.  This enables real-time monitoring in ICS while significantly reducing memory usage and power consumption.

The remainder of the paper is organized as follows: Section \ref{sec:rwork} presents related works in DDDAS and MTD. Section \ref{sec:rml} describes the proposed approach. Section \ref{sec:ear} presents the experimental results. Finally, we conclude the paper in Section \ref{sec:con}.

\section{Related Work}
\label{sec:rwork}
\vspace*{-0.4cm}


In this section, we briefly introduce the DDDAS and MTD. DDDAS (Dynamic Data Driven Applications Systems) has a feedback control loop that integrates an application system's computation and instrumentation aspects dynamically \cite{darema2023handbook}. In this way, the instrumentation data can help to tune the execution, and the execution can control the instrumentataion data in turn. DDDAS-based ICS applications have benefitted a lot from ML recently \cite{darema2005grid}. The authors in~\cite{hong2021dddas} used DDDAS and ML to secure the cyber-environment for the ICS that operates national and military power grids. In~\cite{combita2018dddas}, the authors proposed a DDDAS-based anomaly detection and response approach to secure ICS. In \cite{gokhale2013stochastic}, the authors used DDDAS paradigm and ML to provide dynamic resource management for critical cyber-physical infrastructure. In~\cite{pantopoulou2020data}, the authors emphasized the importance and connection of the DDDAS paradigm and ML to secure cyber-physical systems. However, these ML-based DDDAS are vulnerable to adversarial attacks, and the research for preventing adversarial attacks for DDDAS is largely underexplored in the literature.

Moving Target Defense (MTD) tries to utilize the change across multiple system dimensions to increase uncertainty and complexity for attackers~\cite{ge2014toward}. By dynamically changing the system's configuration, the probability of a successful attack is reduced, and the cost and effort of attacking are significantly increased in turn. The study in \cite{rehman2024proactive} introduces a security framework that dynamically change IoT devices' configuration and location. The work in \cite{tibom2022design} leverages MTD to move the critical applications between virtual and physical nodes.

\section{The Proposed Approach}
\label{sec:rml}
\vspace*{-0.2cm}
This section presents an MTD, TinyML, and DDDAS-based reML approach to combat adversarial attacks. First, the Resilient DDDAS (rDDDAS) development environment is used to design reML for deployment and inference at edge devices. Then, the architecture of reML is described. After that, the process of DAGT is discussed.
\vspace*{-0.2cm}
\subsection{The rDDDAS Development Environment for reML}
\begin{figure}
    \centering
    \includegraphics[width=1\textwidth]{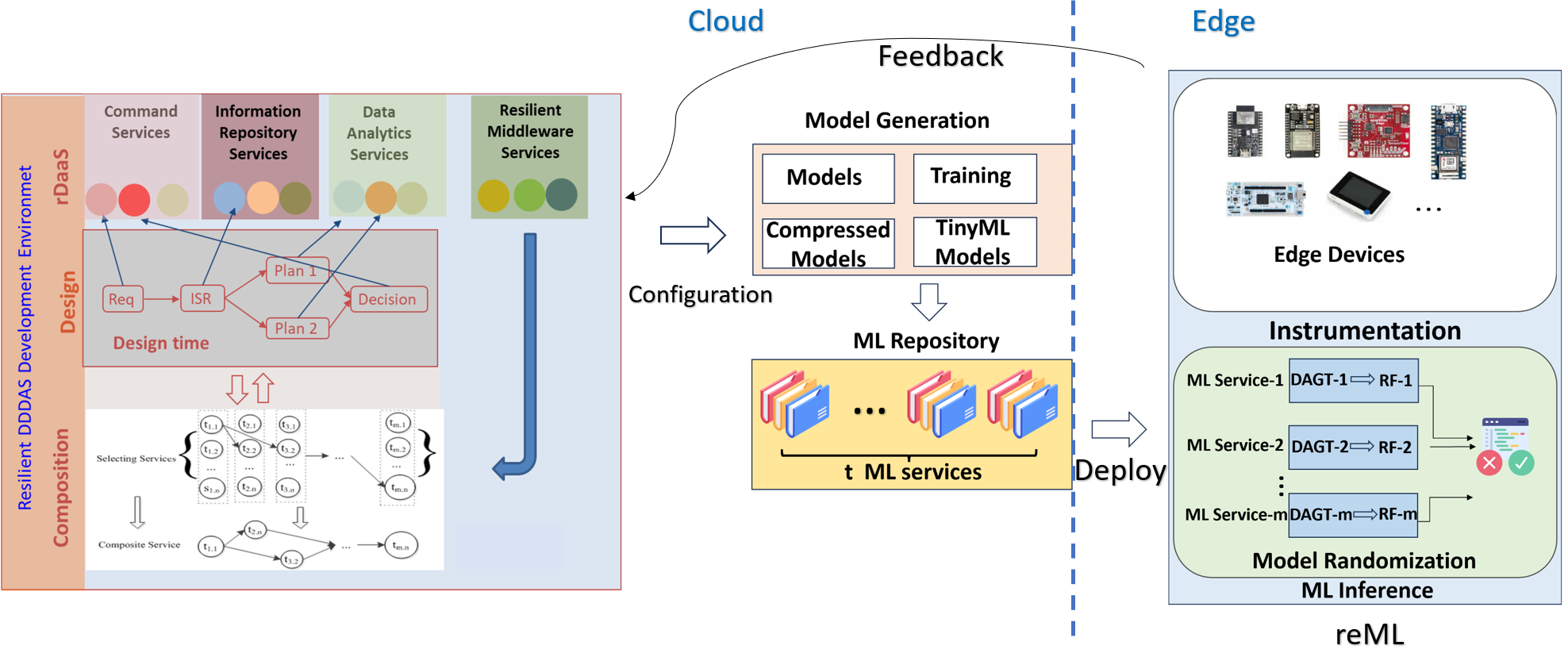}
    \caption{The rDDDAS development architecture for reML.}
    \label{fig:rDDDAS_arch}
\end{figure}

\vspace*{-0.2cm}
 We design the rDDDAS development environment that provides the interface to help users design their reML for edge devices, offer different kinds of services that are needed to construct the system and application, and manage the whole execution and performance, as shown in Figure \ref{fig:rDDDAS_arch}. The environment internally has a collection consisting of different kinds of services, such as command services that handle requests and responses, information repository services that serve static data, and data analytics that support machine learning tasks. During execution, these services will be selected and managed by Resilient Middleware Services. The selection and management are based on the Moving Target Defense (MTD) algorithm. The environment exposes an interface that lists the kinds of services as options and how the services will interact with each other. Users can drag the services to a panel directly and specify the configuration of each service and the way of interaction (Stage 1). We call these services in the design Abstract Services. After submitting the design, Resilient Middleware Services will select Executable Services accordingly, and construct them as a workflow (Stage 2). Based on MTD, the workflow for ML training and model compression is executed (Stage 3). The ML training and model compression are executed and prepared for deployment configuration (Stage 4). The deployment process for executing runtime application services toward lightweight edge ML demand is executed (Stage 5). The edge devices send feedback information, such as prediction, performance measurements, and concept drift detection~\cite{gama2014survey,lopez2023machine},  to the cloud side, leading to dynamic adaptation and ML services updating (Stage 6). The whole process of executable services selection, composition, configuration, deployment, and feedback is processed automatically.
\vspace*{-0.6cm}

\vspace*{-0.8cm}
\subsection{Resilient Edge Machine Learning (reML)}
\vspace*{-0.2cm}
Most of the defense solutions for Adversarial ML attacks focus on input verification, feature extraction and selection methods to enhance security, which are tedious and require extensive time to adapt and change the existing ML algorithms. The reML approach takes a significantly different approach by assuming that attackers will succeed and the goal is to tolerate their attacks in a similar manner to the approach used in fault-tolerant computing. By using MTD, the reML approach changes the diversified ML services at runtime. Diversified ML services mean we randomize ML services for each input instant for detection purposes. Hence, the adversary is not able to misguide the ML pipeline due to not knowing the type of ML models being used at any instant. To support this, reasonably accurate ML models for the same task are selected to construct the ML service ensembles. A user can use rDDDAS development environment to develop the reML for specific edge ML applications. The reML randomly selects and loads $n$ services from $t$ services in the ML repository, and is then deployed to the edge. During the runtime of the reML, the input instance sent to the reML could contain either a clean or an adversarial input. For each input, the reML randomly selects \textbf{$m$} ML services from the loaded {$n$} ML services where each ML service including DAGT for feature space anonymization and random forest (RF) \cite{shao2020ensemble} for classification. Each of the ML services evaluates the input from the user and provides a classification result as the output. Next, a voting mechanism is applied to all the outputs to determine the final output based on the Boyer-Moore majority vote \cite{boyer1991mjrty}.

\begin{figure}
    \centering
    \includegraphics[width=0.8\textwidth]{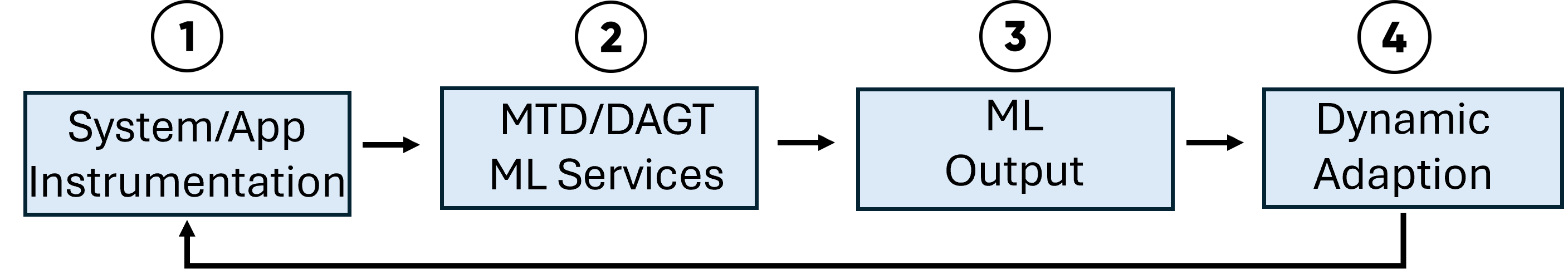}
    \caption{The diagram of reML-based DDDAS model-instrumentation feedback-control loop.}
    \label{fig:rml_arch}
    \vspace*{-1cm}
\end{figure}

Fig.\ref{fig:rml_arch} shows the diagram of reML and how it supports the DDDAS paradigm. In DDDAS methodology, the instrumented data is received continuously from the system/application instrumentation module shown in Step 1. In Step 2, we apply the MTD and DAGT transformation to basically hide the data being used by the randomized ML algorithms to perform their predictions that will be produced by Step 3. The dynamic adaptation of the DDDAS system/application is performed in Step 4 and applied to the system/application in Step 1 using the instrumented sensors and actuators. Without our reML approach, the ML algorithms can be compromised by Adversarial ML attacks. But, by using MTD and DAGT techniques, the attackers will not be able to compromise the ML operations since they have no idea of the data used in the training and prediction because of the DAGT method and also not knowing the ML algorithm used because of the randomization of the ML algorithms.

\vspace*{-0.4cm}

\subsection{Data Air Gap Transformation for Data Feature Space Anonymization}
\begin{figure}
    \centering
    \includegraphics[width=1\textwidth]{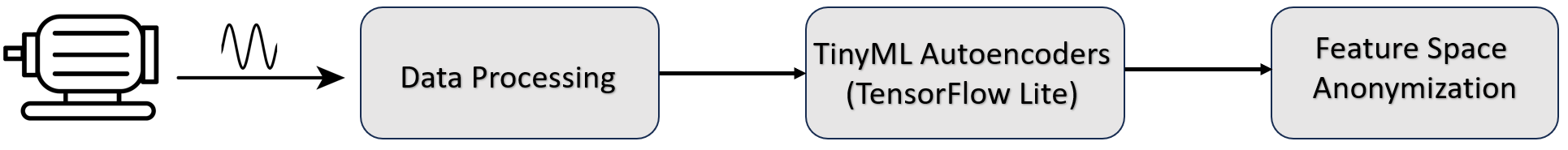}
    \caption{The process of DAGT for feature space anonymization.}
    \label{fig:DAGT}
    \vspace*{-0.8cm}
\end{figure}

\vspace*{-0.2cm}
The core component of reML is Data Air Gap Transformation (DAGT), which uses a TinyML-based deep autoencoder neural network~\cite{zhou2017anomaly} to improve the resiliency of the reML by performing feature space anonymization. We introduced anonymizer algorithms that can randomly transform the original input data points into different data feature spaces by autoencoders that will be used to train the associated RF models. This will make it extremely difficult for the attacker to compromise the ML output because its adversarial ML attacks will be based on the original input data feature space, while the ML models were trained on a completely different data feature space. Anonymizer algorithms based on multi-layer nonlinear feature space transformation through deep autoencoder can be much more difficult to attack than a simple linear transformation, and therefore the effect of adversarial attacks will be minimized. In the rDDDAS environment, the deep autoencoders and RFs are well-trained and model compression is also performed to make them able to execute on resource-constrained edge devices. TensorFlow Lite~\cite{warden2019tinyml} is the leading inference tool designed to compress machine learning models to fit on resource-constrained edge devices. Deep autoencoders are first built using TensorFlow and trained with processed ICS data. Once trained, the deep models are compressed into TinyML models using TensorFlow Lite and thus can be deployed as reML's DAGT that are executed on low-resource edge devices with smaller model sizes and lower inference time while still keeping the representation learning capability of deep models for feature space anonymization. The various settings of hyperparameters of deep autoencoders are applied. Each autoencoder with a different setting is used as feature space anonymization in each ML service. The process of DAGT is shown in Fig.\ref{fig:DAGT}. Tensorflow Decision Forests~\cite{guillame2023yggdrasil} is used to obtain efficient RF models. During runtime, each TinyML-based autoencoder encodes the input ICS data input into the representation of the code layer. The encoded input is then predicted by an RF (Random Forest) model trained on the training dataset processed by the TinyML-based autoencoder.

\vspace*{-0.6cm}
\vspace*{-0.3cm}

\vspace*{0.1cm}
\section{Experimental Results}
\label{sec:ear}
\vspace*{-0.1cm}
\subsection{Experimental Setup}
\vspace*{-0.1cm}
The ICS datasets developed by Mississippi State University and Oak Ridge National Laboratory implement a scaled-down version of a power system framework, which is used to conduct the experiments~\cite{anthi2021adversarial,upadhyay2021intrusion,pan2015classification}. The datasets have both benign and malicious data points generated from the power system. These data points have been further categorized into three main classes; ‘no event’ instances, ‘natural event’ instances, and ‘attack event’ instances. Both the ‘no event’ and ‘natural event’ instances are grouped together to represent benign activity in this study. To generate the malicious data, attacks from five scenarios were deployed on the power system, including short-circuit fault, line maintenance, remote tripping command injection, relay setting change attack, and data injection attack. The popular Jacobian-based Saliency Map Attack (JSMA) \cite{papernot2016limitations} is used to generate adversarial test samples. We use JSMA in a black-box setting. The training data is used for training, and the clean and adversarial test data is used to evaluate methods. The base model and our previous rML(resilient ML) approach~\cite{yao2023resilient} are used to compare with the proposed reML. RF is used as the base model. The difference between rML and reML is that rML keeps the original larger model without applying TinyML for model compression. Since we only consider the black-box attack of JSMA, a reasonable assumption of the ICS scenario is that an adversarial attack is launched after the input sample is loaded into rML and reML, and only minority ML services are affected by adversarial data thanks to the architecture of rML and reML. $F_1\text{-Score}$, $Precision$, and $Recall$ are used as the metrics for detection performance evaluations on the test data. 
\vspace*{-0.3cm}
\subsection{Results}
\vspace*{-0.1cm}

\begin{figure*}[htb]
\vspace*{-0.5cm}
    \centering
    \includegraphics[width=0.8\textwidth]{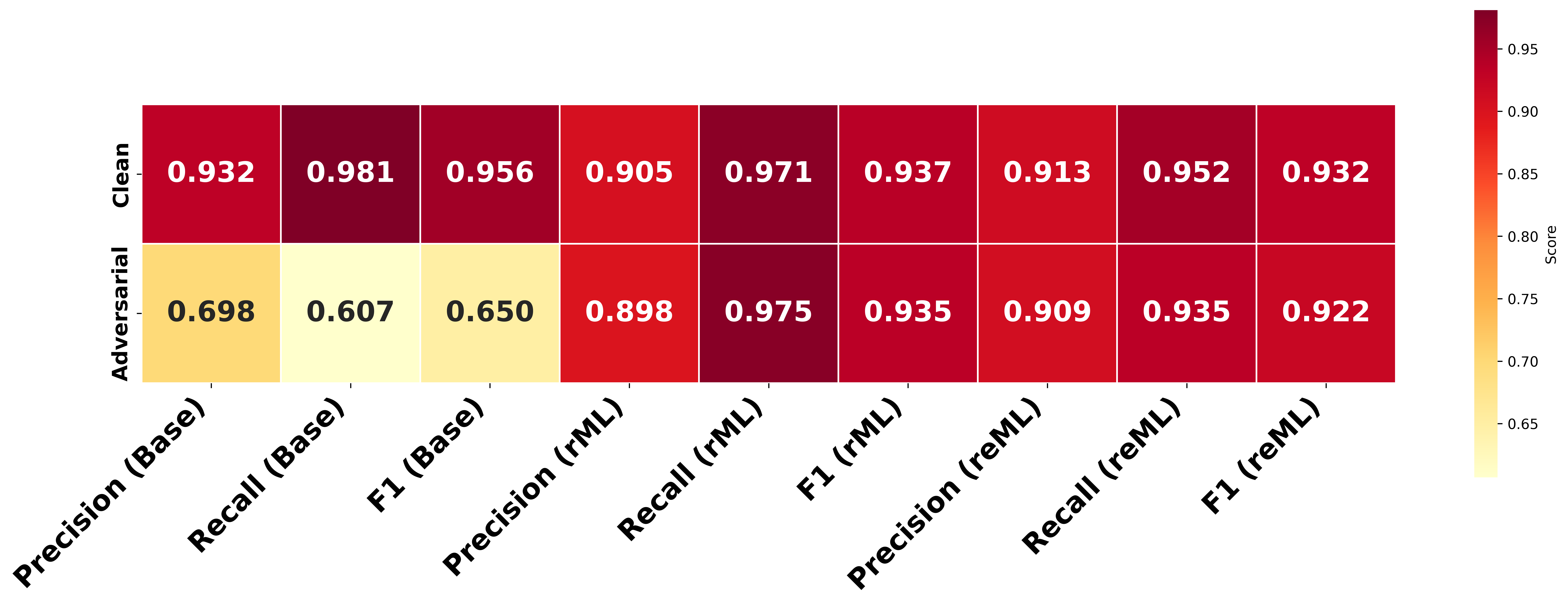}
    \caption{The detection performance heatmap for comparing different methods.}
    \label{fig:performance_heatmap}
    \vspace*{-0.6cm}
\end{figure*}

\begin{figure*}[htb!]
    \vspace*{-0.4cm}
    \centering
    \includegraphics[width=0.8\textwidth]
    {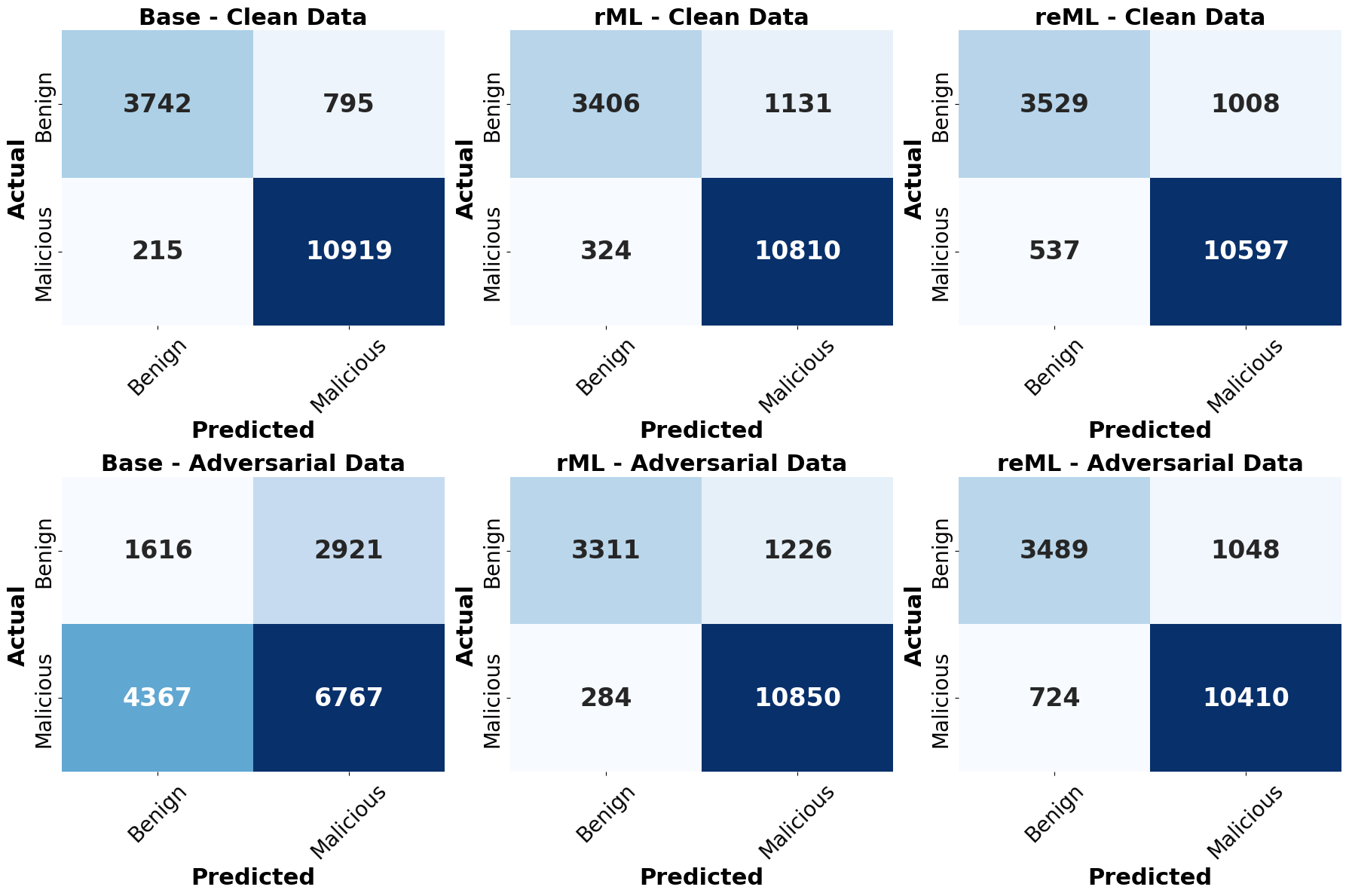}
    \caption{Confusion matrix on clean and adversarial test dataset with different methods.}
    \label{fig:confusion_matrix}
    \vspace*{-0.4cm}
\end{figure*}

Figure \ref{fig:performance_heatmap} shows the performance heatmap on clean and adversarial test data for comparing different methods, including the base model (random forest), the rML, and the reML. Figure \ref{fig:confusion_matrix} shows the confusion matrix on clean and adversarial test data using different Methods. As we can see from these two figures, the base method (RF) performs well on clean test data but fails significantly on adversarial test data, demonstrating the impact of the adversarial attacks on the base model. The rML approach shows a significant improvement method over the base method on adversarial test data and achieves similar performance on clean test data. However, the rML cannot be deployed to resource-constrained devices that cannot provide the required computational resources. In comparison with the rML, reML has only a very slight drop in prediction capability and performs well on both clean and adversarial test data. The results validate the feasibility of reML and its robustness against adversarial attacks due to its resilience features like DAGT and model randomization. Importantly, reML uses TinyML and thus can run on resource-constrained edge devices and keep data on the edge to preserve data privacy.
\vspace*{-0.3cm}

\vspace*{-0.3cm}
\section{Conclusion}
\vspace*{-0.4cm}
In this paper, we presented a resilient edge machine learning (reML) architecture designed to safeguard industrial control systems (ICS) against adversarial attacks. By leveraging the Dynamic Data-Driven Applications Systems (DDDAS) paradigm, Moving Target Defense (MTD) theory, and TinyML framework, reML demonstrates robust defense mechanisms through Data Air Gap Transformation (DAGT) and model randomization. Our experimental results validate the efficacy of reML in detecting and mitigating adversarial threats while maintaining low-latency and power-efficient operations on resource-constrained edge devices.
The reML architecture's ability to perform on-device attack detection and local ML inference significantly enhances ICS security by reducing response times and preserving data privacy. Our study validates that reML provides a viable and effective solution for implementing resilient machine learning at the edge devices, ensuring the security and reliability of critical infrastructure systems.

While reML aims to provide resilience against adversarial attacks, scaling this approach to large-scale ICS environments could present challenges. Hence, future research will focus on optimizing the approach for large-scale deployments, reducing computational overheads, improving compatibility with existing ICS infrastructures, expanding the resilience capabilities against a broader range of adversarial attacks, and using more advanced autoencoders to feature anonymization.

\label{sec:con}

\vspace*{-0.4cm}
\section{Acknowledgment}
\vspace*{-0.4cm}
This work is supported by the National Science Foundation (NSF)
projects 1624668, 1921485, OIA-2218046, the Department of Energy/National Nuclear Security Administration under Award Number
DE-NA0003946, and the AGILITY project 4263090, sponsored by Korea
Institute for Advancement of Technology.
\vspace*{-0.4cm}
\bibliographystyle{IEEEtran}
\vspace*{-0.4cm}
\bibliography{refs}

\begin{thebibliography}{10}
\providecommand{\url}[1]{#1}
\csname url@samestyle\endcsname
\providecommand{\newblock}{\relax}
\providecommand{\bibinfo}[2]{#2}
\providecommand{\BIBentrySTDinterwordspacing}{\spaceskip=0pt\relax}
\providecommand{\BIBentryALTinterwordstretchfactor}{4}
\providecommand{\BIBentryALTinterwordspacing}{\spaceskip=\fontdimen2\font plus
\BIBentryALTinterwordstretchfactor\fontdimen3\font minus \fontdimen4\font\relax}
\providecommand{\BIBforeignlanguage}[2]{{%
\expandafter\ifx\csname l@#1\endcsname\relax
\typeout{** WARNING: IEEEtran.bst: No hyphenation pattern has been}%
\typeout{** loaded for the language `#1'. Using the pattern for}%
\typeout{** the default language instead.}%
\else
\language=\csname l@#1\endcsname
\fi
#2}}
\providecommand{\BIBdecl}{\relax}
\BIBdecl

\bibitem{chen2020generating}
J.~Chen, X.~Gao, R.~Deng, Y.~He, C.~Fang, and P.~Cheng, ``Generating adversarial examples against machine learning-based intrusion detector in industrial control systems,'' \emph{IEEE Transactions on Dependable and Secure Computing}, 2020.

\bibitem{darema2005grid}
F.~Darema, ``Grid computing and beyond: The context of dynamic data driven applications systems,'' \emph{Proceedings of the IEEE}, vol.~93, no.~3, pp. 692--697, 2005.

\bibitem{papernot2016limitations}
N.~Papernot, P.~McDaniel, S.~Jha, M.~Fredrikson, Z.~B. Celik, and A.~Swami, ``The limitations of deep learning in adversarial settings,'' in \emph{2016 IEEE European symposium on security and privacy (EuroS\&P)}.\hskip 1em plus 0.5em minus 0.4em\relax IEEE, 2016, pp. 372--387.

\bibitem{yao2023resilient}
L.~Yao, S.~Shao, and S.~Hariri, ``Resilient machine learning (rml) against adversarial attacks on industrial control systems,'' in \emph{2023 20th ACS/IEEE International Conference on Computer Systems and Applications (AICCSA)}.\hskip 1em plus 0.5em minus 0.4em\relax IEEE, 2023.

\bibitem{blasch2019study}
E.~Blasch, R.~Xu, S.~Y. Nikouei, and Y.~Chen, ``A study of lightweight dddas architecture for real-time public safety applications through hybrid simulation,'' in \emph{2019 Winter Simulation Conference (WSC)}.\hskip 1em plus 0.5em minus 0.4em\relax IEEE, 2019, pp. 762--773.

\bibitem{ge2014toward}
L.~Ge, W.~Yu, D.~Shen, G.~Chen, K.~Pham, E.~Blasch, and C.~Lu, ``Toward effectiveness and agility of network security situational awareness using moving target defense (mtd),'' in \emph{Sensors and Systems for Space Applications VII}, vol. 9085.\hskip 1em plus 0.5em minus 0.4em\relax SPIE, 2014, pp. 185--193.

\bibitem{warden2019tinyml}
P.~Warden and D.~Situnayake, \emph{Tinyml: Machine learning with tensorflow lite on arduino and ultra-low-power microcontrollers}.\hskip 1em plus 0.5em minus 0.4em\relax O'Reilly Media, 2019.

\bibitem{darema2023handbook}
F.~Darema, E.~P. Blasch, S.~Ravela, and A.~J. Aved, \emph{Handbook of Dynamic Data Driven Applications Systems: Volume 2}.\hskip 1em plus 0.5em minus 0.4em\relax Springer Nature, 2023.

\bibitem{hong2021dddas}
Y.~Hong and I.~I.~O. TECHNOLOGY, ``Dddas-crafts: A dddas-based cyber-resilient and attack-secure framework for trustworthy industrial control systems,'' 2021.

\bibitem{combita2018dddas}
L.~F. Combita, J.~A. Giraldo, A.~A. Cardenas, and N.~Quijano, ``Dddas for attack detection and isolation of control systems,'' \emph{Handbook of Dynamic Data Driven Applications Systems}, pp. 407--422, 2018.

\bibitem{gokhale2013stochastic}
A.~Gokhale, X.~Koutsoukos, and D.~Schmidt, ``Stochastic hybrid systems modeling and middleware-enabled dddas for next-generation us air force systems,'' \emph{aFOSR DDDAS-funded project (\# FA9550-13-1-0227}, 2013.

\bibitem{pantopoulou2020data}
S.~Pantopoulou, P.~L. Lagari, C.~H. Townsend, and L.~H. Tsoukalas, ``Data-based defense-in-depth of critical systems,'' in \emph{Dynamic Data Driven Applications Systems: Third International Conference, DDDAS 2020, Boston, MA, USA, October 2-4, 2020, Proceedings 3}.\hskip 1em plus 0.5em minus 0.4em\relax Springer, 2020, pp. 283--290.

\bibitem{rehman2024proactive}
Z.~Rehman, I.~Gondal, M.~Ge, H.~Dong, M.~Gregory, and Z.~Tari, ``Proactive defense mechanism: Enhancing iot security through diversity-based moving target defense and cyber deception,'' \emph{Computers \& Security}, vol. 139, p. 103685, 2024.

\bibitem{tibom2022design}
P.~Tibom and M.~Buck, ``Design, implementation and evaluation of a moving target defense in distributed systems,'' 2022.

\bibitem{gama2014survey}
J.~Gama, I.~{\v{Z}}liobait{\.e}, A.~Bifet, M.~Pechenizkiy, and A.~Bouchachia, ``A survey on concept drift adaptation,'' \emph{ACM computing surveys (CSUR)}, vol.~46, no.~4, pp. 1--37, 2014.

\bibitem{lopez2023machine}
M.~M. Lopez, S.~Shao, S.~Hariri, and S.~Salehi, ``Machine learning for intrusion detection: Stream classification guided by clustering for sustainable security in iot,'' in \emph{Proceedings of the Great Lakes Symposium on VLSI 2023}, 2023, pp. 691--696.

\bibitem{shao2020ensemble}
S.~Shao, C.~Tunc, A.~Al-Shawi, and S.~Hariri, ``An ensemble of ensembles approach to author attribution for internet relay chat forensics,'' \emph{ACM Transactions on Management Information Systems (TMIS)}, vol.~11, no.~4, pp. 1--25, 2020.

\bibitem{boyer1991mjrty}
R.~S. Boyer and J.~S. Moore, ``Mjrty—a fast majority vote algorithm,'' in \emph{Automated reasoning: essays in honor of Woody Bledsoe}.\hskip 1em plus 0.5em minus 0.4em\relax Springer, 1991, pp. 105--117.

\bibitem{zhou2017anomaly}
C.~Zhou and R.~C. Paffenroth, ``Anomaly detection with robust deep autoencoders,'' in \emph{Proceedings of the 23rd ACM SIGKDD international conference on knowledge discovery and data mining}, 2017.

\bibitem{guillame2023yggdrasil}
M.~Guillame-Bert, S.~Bruch, R.~Stotz, and J.~Pfeifer, ``Yggdrasil decision forests: A fast and extensible decision forests library,'' in \emph{Proceedings of the 29th ACM SIGKDD Conference on Knowledge Discovery and Data Mining}, 2023.

\bibitem{anthi2021adversarial}
E.~Anthi, L.~Williams, M.~Rhode, P.~Burnap, and A.~Wedgbury, ``Adversarial attacks on machine learning cybersecurity defences in industrial control systems,'' \emph{Journal of Information Security and Applications}, vol.~58, p. 102717, 2021.

\bibitem{upadhyay2021intrusion}
D.~Upadhyay, J.~Manero, M.~Zaman, and S.~Sampalli, ``Intrusion detection in scada based power grids: Recursive feature elimination model with majority vote ensemble algorithm,'' \emph{IEEE Transactions on Network Science and Engineering}, vol.~8, no.~3, pp. 2559--2574, 2021.

\bibitem{pan2015classification}
S.~Pan, T.~Morris, and U.~Adhikari, ``Classification of disturbances and cyber-attacks in power systems using heterogeneous time-synchronized data,'' \emph{IEEE Transactions on Industrial Informatics}, vol.~11, no.~3, pp. 650--662, 2015.

\end{thebibliography}
\vspace*{-0.8cm}
\end{document}